\documentclass[pra,twocolumn,aps,superscriptaddress,showpacs,floatfix]{revtex4}
\usepackage{amssymb}
\usepackage{graphicx}
\usepackage{dcolumn}
\usepackage{bm}
\usepackage{amsmath}
\usepackage{amsbsy}
\usepackage{amssymb}
\usepackage{graphicx}
\usepackage{color}
\usepackage{epstopdf}
\usepackage{mathrsfs}
\usepackage{multirow}
\usepackage{ulem}
\usepackage{comment}

\usepackage[colorlinks,linkcolor=magenta,citecolor=blue,urlcolor=blue]{hyperref}
\usepackage{etoolbox}
\makeatletter
\patchcmd{\@makecaption}
{\scshape}
{}
{}
{}
\makeatother

\makeatletter
\newcommand{\Rmnum}[1]{\expandafter\@slowromancap\romannumeral #1@}
\makeatother

\begin{document}

	\title{Size-Dependent Skin Effect Transitions in Weakly Coupled Non-Reciprocal Chains}
	%%%%%%%%authors%%%%%%%%%%%%%%%
	\author{Yixuan Li}
	\affiliation{Institute of Theoretical Physics and State Key Laboratory of Quantum Optics Technologies and Devices, Shanxi University, Taiyuan 030006, China}	
	
	\author{Linhu Li}
	\email{lilinhu@quantumsc.cn}
	\affiliation{Quantum Science Center of Guangdong-Hong Kong-Macao Greater Bay Area (Guangdong), Shenzhen, China}
	
	\author{Zhihao Xu}
	\email{xuzhihao@sxu.edu.cn}
	\affiliation{Institute of Theoretical Physics and State Key Laboratory of Quantum Optics Technologies and Devices, Shanxi University, Taiyuan 030006, China}
	\affiliation{Collaborative Innovation Center of Extreme Optics, Shanxi University, Taiyuan 030006, China}
	%%%%%%%%authors%%%%%%%%%%%%%%%
	
	\date{\today}
	
	\begin{abstract}
		Non-Hermitian systems exhibit unique boundary phenomena absent in their Hermitian counterparts, most notably the non-Hermitian skin effect (NHSE). In this work, we explore a lattice model consisting of two coupled non-reciprocal chains, focusing on the interplay between system size, inter-chain coupling, and spectral topology. Using both analytical and numerical approaches, we systematically examine the evolution of the complex energy spectra and spectral winding numbers under periodic and open boundary conditions. Our results uncover a variety of size-dependent localization transitions, including the emergence and instability of concurrent bipolar skin effects in the $W=0$ region, and their crossover to unipolar and conventional bipolar NHSE as the system size increases. Notably, we demonstrate that these size-dependent behaviors persist even beyond the weak-coupling regime, highlighting their universality in non-Hermitian systems with complex spectral structures. This study provides insights into the mechanisms governing skin effects and offers practical guidelines for engineering non-Hermitian topological phases in synthetic lattices.		
	\end{abstract}
	\maketitle

	\section{Introduction}
	
	Non-Hermitian systems have recently emerged as a fertile platform for discovering novel physical phenomena that are forbidden in conventional Hermitian settings~\cite{Shen2018,Ashida2020,Bergholtz2021,Bender1998,Borgnia2020,Kunst2018,Song2019,Guo2009,BB2025}. Among these, the non-Hermitian skin effect (NHSE)~\cite{Yao2018,2Yao2018,Lee2019,Helbig2020,Xiao2020,Weidemann2020,Longhi2019,Liang2022,Yi2020,Longhi2022,Longhi2022B,Zeng2022,2Zeng2022,Zhang2022,Lai2025,Li2025,Hu2025,FengMei2025}--the accumulation of a macroscopic number of eigenstates at the boundaries--has attracted considerable attention due to its fundamental significance and wide-ranging applications in photonics, electronics, and acoustics~\cite{Ghatak2020,Gou2020,Li2020,Yoshida2020,Mandal2020,Gao2020,Zhu2020,Hofmann2020,Brandenbourger2019,Zhong2021,2Zhang2021,Wang2025,Pu2025}. The NHSE is intimately linked to the topology of the complex energy spectrum, which can be characterized by the spectral winding number~\cite{Zhang2020,Okuma2020,Borgnia2020,Gong2018,Kawabata2019}. This topological invariant not only signals the presence or absence of NHSE, but also distinguishes between different types of skin modes, such as unipolar and bipolar skin effects~\cite{2Zhang2021,Rafi-UI-Islam2024,Zhao2024}.
	
	Recently, the fragility of NHSE has been uncovered, revealing that skin localization under weak perturbations exhibits size-dependent features and may ultimately vanish in the thermodynamic limit~\cite{li2020nat,guo2021exact,liu2021exact,li2021impurity,Guo2023}. In particular, a novel type of critical NHSE has been identified in systems where multiple non-Hermitian channels with distinct skin localizations are weakly coupled~\cite{li2020nat,Yokomizo2021B,2Yokomizo2021B,Liang2022CPB,Rafi-UI-Islam2022,Qin2023,Xu2025,Liu2020,Yang2024,Liu2024,Li2025concurrent}. The hallmark of the critical NHSE is a discontinuous jump in both the energy spectrum and the spatial distribution of eigenstates when the system size exceeds a critical threshold~\cite{li2020nat,Yokomizo2021B}. In the critical regime, each eigenstate under open boundary conditions (OBCs) becomes concurrently localized at different ends of the system, and the OBC spectrum cannot be continuously connected to that of the decoupled system, reflecting a fundamental noncommutativity between the thermodynamic and zero-coupling limits~\cite{li2020nat}. Recent theoretical advances have further extended the concept of the critical NHSE to many-body and higher-dimensional systems, uncovering universal scaling laws and critical exponents that govern the crossover between different localization regimes~\cite{Qin2025,Gliozzi2024,Hamazaki2019,Mukherjee2021,Moudgalya2022,Fromholz2020,liang2025intrinsic,ou2025anisotropic}. 
	
	In this work, we systematically study a lattice model composed of two weakly coupled nonreciprocal chains and uncover a rich variety of size-dependent transitions---from a concurrent bipolar skin effect (CBSE) at finite sizes to unipolar NHSE or conventional bipolar NHSE in the thermodynamic limit. Notably, the CBSE identified here is distinct from conventional bipolar NHSE, where left- and right localization appear in different eigenstates whose eigenenergies carry positive and negative topological invariant‌s, respectively~\cite{2Zhang2021,Rafi-UI-Islam2024,Zhao2024}. The various skin localizations and their transitions can be topologically characterized by the spectral winding number. Specifically, we find that CBSE emerges for finite-size states whose eigenenergies have a zero winding number, but these states eventually evolve into regions supporting nonzero winding numbers and unipolar NHSE or conventional bipolar NHSE as the system size increases. Furthermore, this size-dependent critical behavior persists even under stronger interchain coupling, rather than only in the limit of vanishing coupling, thus facilitating experimental realization by using electrical circuits. Our findings not only deepen our understanding of non-Hermitian criticality and the interplay among nonreciprocity, topology, and boundary effects, but also provide guidance for engineering NHSE in non-Hermitian systems, with potential applications in future topological devices.
	
	\section{Model and Hamiltonian}
	
	\begin{figure}[htbp]
		\includegraphics[width=0.5\textwidth]{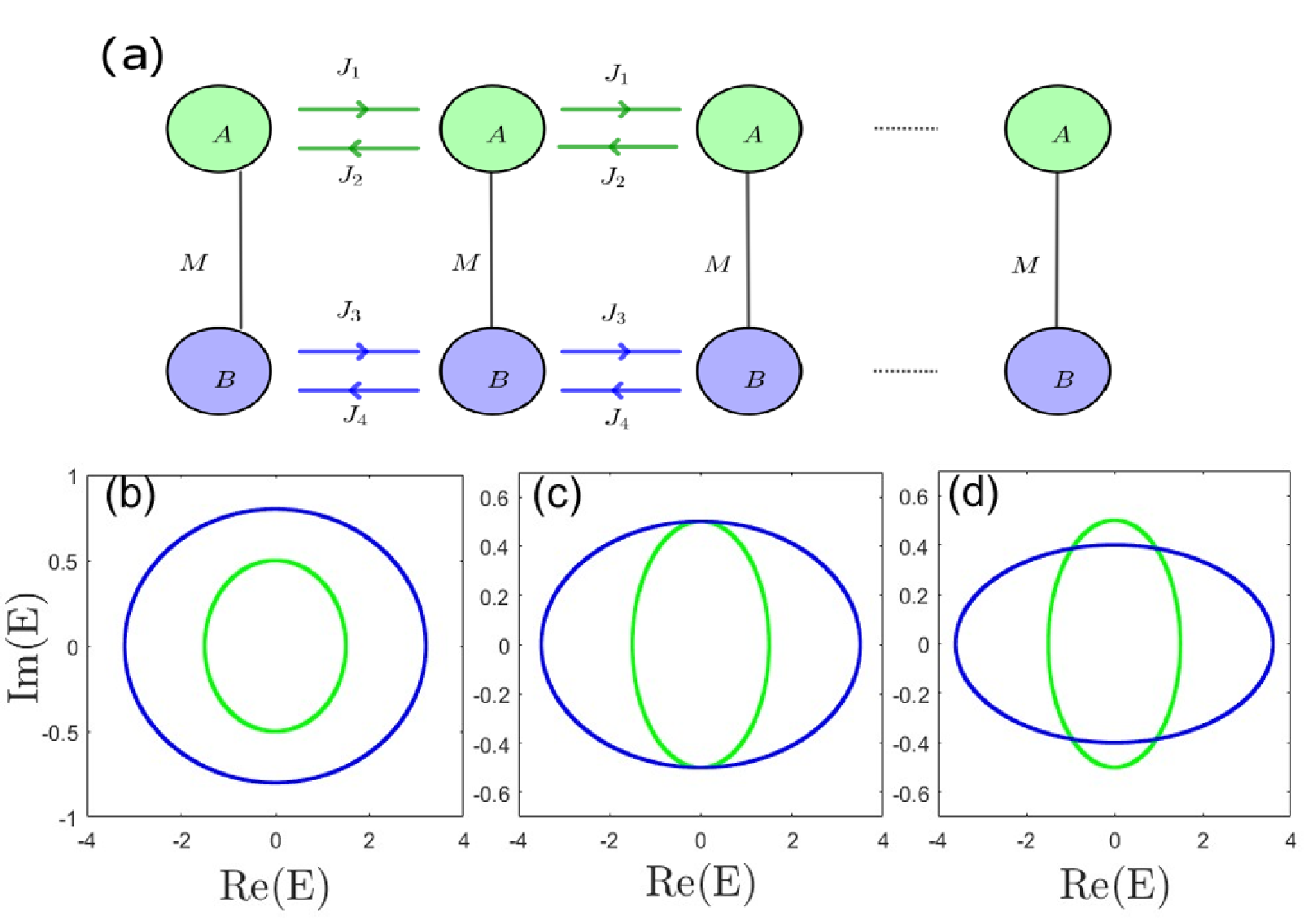}
		\caption{(a) Schematic illustration of two weakly coupled nonreciprocal chains, labeled $A$ and $B$, with couple amplitude $M$. The right- and left-directed hopping amplitudes of the $A$ chain are $J_1$ and $J_2$, respectively, while $J_3$ and $J_4$ denote the right- and left-directed hopping amplitudes of the $B$ chain. Energy spectra of the two decoupled chains under periodic boundary conditions (PBCs) for (b) $\delta_b=0.8$, (c) $\delta_b=0.5$, and (d) $\delta_b=0.4$. Here, $J_1=0.5$, $J_3=2$, $\delta_a=0.5$, and $M=0$.}
		\label{Fig1}
	\end{figure}
	
	We consider two weakly coupled, nonreciprocal chains, labeled $A$ and $B$, respectively. The Hamiltonian of the system is given by  
	\begin{equation} 
		\begin{aligned}
			\hat{H} = & \sum_{j} \left( J_1 \hat{a}_{j + 1}^{\dagger} \hat{a}_{j}+J_2 \hat{a}_{j}^{\dagger} \hat{a}_{j + 1} + J_3 \hat{b}_{j + 1}^{\dagger} \hat{b}_{j} + J_4 \hat{b}_{j}^{\dagger} \hat{b}_{j + 1} \right) \\
			& + \sum_{j} M \left( \hat{a}_{j}^{\dagger} \hat{b}_{j} + \hat{a}_{j} \hat{b}_{j}^{\dagger} \right) \label{eq1},
		\end{aligned}
	\end{equation}  
	where $\hat{a}^{\dagger}_{j}$ ($\hat{a}_{j}$) and $\hat{b}^{\dagger}_{j}$ ($\hat{b}_{j}$) are the creation (annihilation) operators for the $A$ and $B$ chains, respectively. This system consists of $N$ unit cells, each containing two sublattices, resulting in a total system size of $2N$. As illustrated in Fig. \ref{Fig1}(a), the nonreciprocal hopping amplitudes for the $A$-chain are $J_1$ and $J_2=J_1 + \delta_a$, while those for the $B$ chain are $J_3$ and $J_4=J_3 - \delta_b$. Here, $\delta_a$ and $\delta_b$ quantify the degrees of nonreciprocity in the corresponding chain. The two chains are coupled via a weak interchain coupling amplitude $M$. Without loss of generality, we assume that $J_1$, $J_4$, $\delta_a$ and $\delta_b$ are nonzero positive real numbers, and focus on the case where $J_3+J_4>J_1+J_2$ (a discussion of the case $J_3+J_4<J_1+J_2$ is provided in Appendix A). 
	
	Under periodic boundary conditions (PBCs), the energy spectrum is given by $E(k) = \left[\mathcal{A}(k) \pm \sqrt{\mathcal{B}(k)}\right]/2$, where $\mathcal{A}(k)=e^{ik}(J_2 + J_4) + e^{-ik}(J_1 + J_3)$, $\mathcal{B}(k)= 2(J_4 - J_2)(J_3 - J_1) + (J_2 - J_4)^2 e^{2ik}+ (J_3 - J_1)^2 e^{-2ik} + 4M^2$, and $k$ is the wave vector. In the absence of interchain coupling ($M=0$), the energy spectra of the two decoupled chains form two independent closed loops (ellipses)  in the complex plane, as shown in Figs. \ref{Fig1}(b) and -\ref{Fig1}(c) for $J_1=0.5$, $J_3=2$, and $\delta_a=0.5$. Here, the green (blue) closed loops corresponds to the spectrum of the $A$ chain ($B$ chain). Since $J_3+J_4>J_1+J_2$, the major axis of the ellipse for the $B$ chain along the real axis of the spectrum is always larger than that of the $A$ chain. The geometric relation between the two complex spectra depends on the specific values of $\delta_a$ and $\delta_b$. As shown in Figs. \ref{Fig1}(b)-\ref{Fig1}(d), when $\delta_a<\delta_b$, the ellipse corresponding to the $A$ chain is entirely enclosed within that of the $B$ chain. For $\delta_a=\delta_b$, the two independent ellipses touch at two points, $\tilde{E}_{1}=(0,-i\delta_a)$ and $\tilde{E}_{2}=(0,i\delta_a)$, in the complex energy plane. When $\delta_b>\delta_a$, the two ellipses intersect, resulting in an overlap of their spectra.
	
	To characterize the topological features in the complex spectral space, one can calculate the winding number, defined as 
	\begin{equation}
		W = \frac{1}{2\pi i} \int_{0}^{2\pi} \mathrm{d}k \; \partial_{k} \arg{\det[\hat{H}(k) - E_{0}]} \label{eq2},
	\end{equation}  
	where $E_{0}$ is a chosen reference energy; $\arg[\cdot]$ denotes the argument of a complex number. In our numerical calculations, we keep the Hamiltonian fixed and vary the position of the reference energy $E_0$ across the complex energy plane. The winding number $W$ counts the number of times the complex spectral trajectory winds around the base energy $E_0$ as the momentum $k$ varies from $0$ to $2\pi$, thereby capturing the emergence of the NHSE \cite{Zhang2020,Okuma2020,Borgnia2020}. Specifically, $W = -1$ indicates a right-directed NHSE, where all wave functions are localized at the right boundary, while $W = +1$ corresponds a left-directed NHSE, with all wave functions localized at the left boundary. For $W=0$, the unipolar NHSE is absent. In the decoupling case with $M=0$, the system exhibits a winding number $W=+1$ for the $A$ chain and $W=-1$ for the $B$ chain. Under OBCs, this leads to left-directed and right-directed NHSEs along the respective chains \cite{Okuma2020,Zhang2020,Gong2018,Kawabata2019}. Consequently, the full system realizes a standard bipolar skin configuration composed of two independent unipolar skin chains. The corresponding OBC spectrum remains entirely real, and no critical NHSE occurs.
	
	In the following, we introduce a weak interchain coupling and systematically investigate its effect on the size-dependent transitions in three representative cases: $\delta_b>\delta_a$, $\delta_a=\delta_b$, and $\delta_a>\delta_b$. This allows us to elucidate how the interplay between nonreciprocity and interchain coupling gives rise to distinct localization behaviors as the system size changes. For clarity, we first focus on the case with a weak interchain coupling $M=0.01$, where the PBC spectrum displays geometric features similar to those in the decoupled limit. We then discuss the NHSE behavior for finite values of $M$ that are comparable to other parameters. Without loss of generality, we set $J_1=0.5$, $J_3=2$, and $\delta_a=0.5$ throughout this paper.
	
	\section{Size-dependent skin transitions in weak coupling limits}

	\subsection{Nested loops with $\delta_{b}>\delta_{a}$}
	
	\begin{figure}[t]  
		\includegraphics[width=0.5\textwidth]{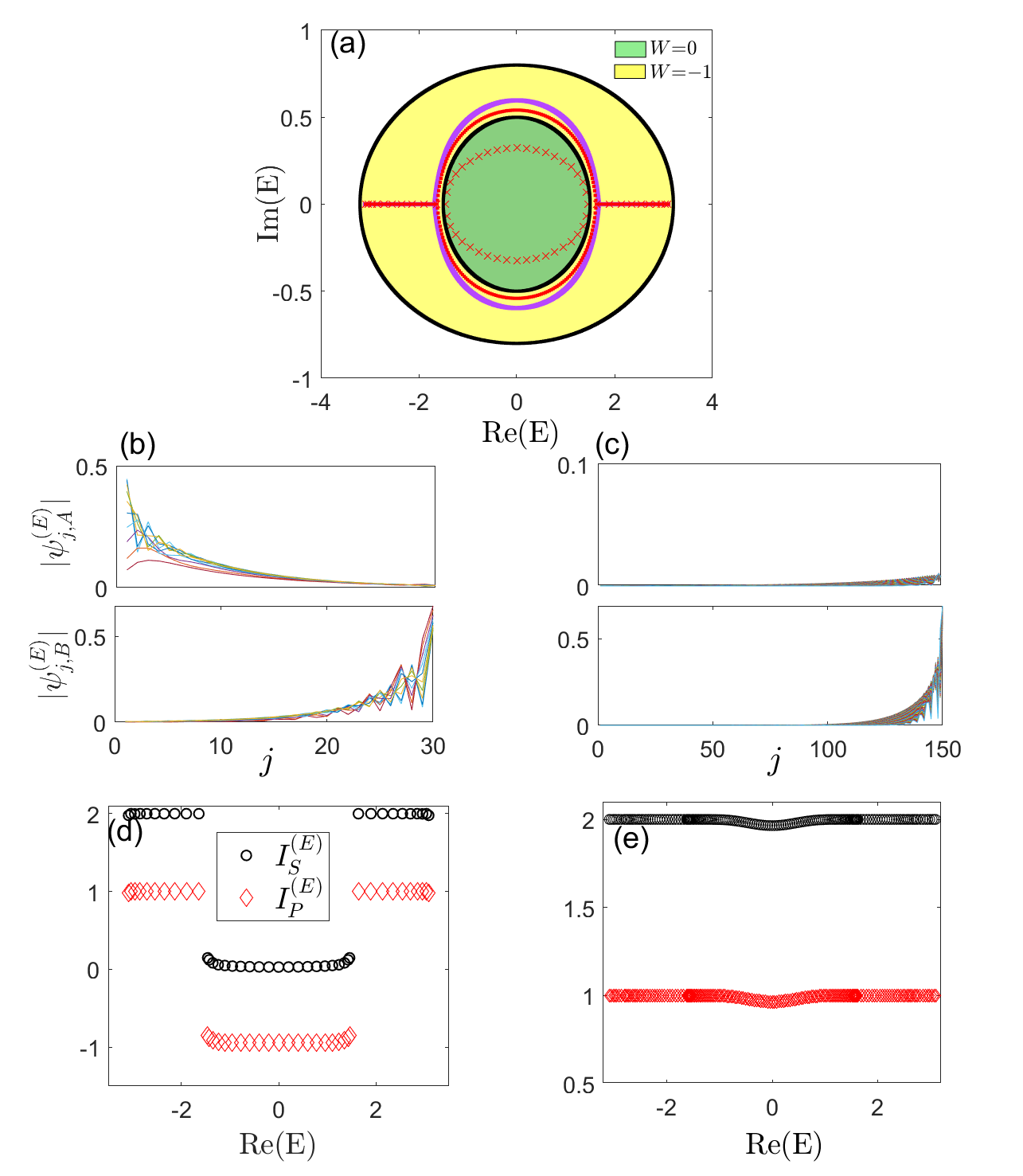}  
		\caption{(a) Energy spectra for $\delta_{b}>\delta_{a}$ under PBCs and OBCs. Black curves represent the spectrum under PBCs, red crosses denote the OBC spectrum for $N = 30$, red dots correspond to $N = 150$, and purple dots indicate the thermodynamic-limit spectrum obtained from the non-Bloch band theory.	The yellow (green) shading marks the $W = -1$ ($W = 0$) region. The spatial profiles $|\psi_{j,A}^{(E)}|$ and $|\psi_{j,B}^{(E)}|$ of all wave functions with eigenvalue $E$ from the complex energy loop under OBCs for (b) $N = 30$ and (c) $N = 150$. The quantities $I_S^{(E)}$ and $I_P^{(E)}$ for different eigenvalues for (d) $N = 30$ and (e) $N = 150$. Here, $J_1 = 0.5$, $J_3 = 2$, $\delta_a = 0.5$, $\delta_b = 0.8 $, and $M = 0.01$.}  
		\label{Fig2}  
	\end{figure}
	
	\begin{figure}[htbp]
		\includegraphics[width=0.5\textwidth]{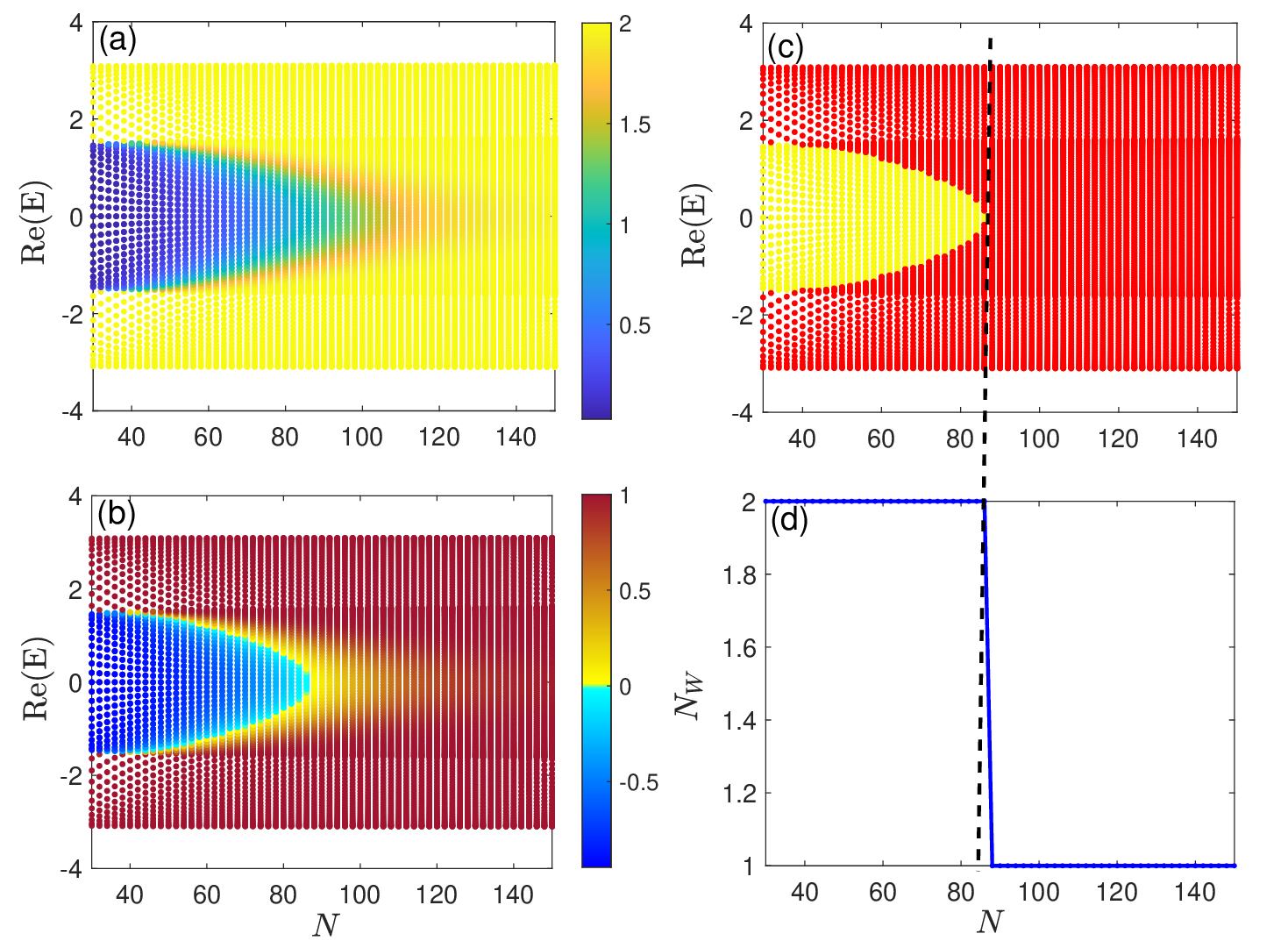}
		\caption{(a) $I_S^{(E)}$ and (b) $I_P^{(E)}$ as functions of $\mathrm{Re}(E)$ and system size $N$. (c) The values of $W_{\mathrm{OBC}}^{(E)}$ for OBC eigenstates and different system size $N$. Yellow dots mark the eigenenergies with $W_{\mathrm{OBC}}^{(E)}=0$, and red dots indicate the eigenenergies with $W_{\mathrm{OBC}}^{(E)}=-1$. (d) The number $N_W$ of different values of the winding number $W_{\mathrm{OBC}}^{(E)}$ as a function of system size $N$. The dashed lines in both (c) and (d) correspond to $N = 86$. Parameters are $J_1 = 0.5$, $J_3 = 2$, $\delta_a = 0.5$, $\delta_b = 0.8$, and $M = 0.01$.
		}
		\label{Fig3}
	\end{figure} 
	
	We first consider the scenario with $\delta_b>\delta_a$, where the two spectral loops under PBCs are nested within each other, as shown in Fig. \ref{Fig2}(a). Using Eq.~(\ref{eq2}), we determine the spectral winding number for reference energies $E_0$ enclosed by the PBC spectrum, with $W=0$ for the region enclosed by both loops (green shaded region) and $W=-1$ for the region enclosed by only a single loop (yellow shaded region).
	
	In contrast to the decoupled case, where all energies are real under OBCs, the energy spectrum here exhibits markedly different characteristics in various regions. For small system sizes, taking $N=30$ as a concrete example in Fig.~\ref{Fig2}(a), all energies within the $W=-1$ region remain real, while the remaining states form a complex loop confined to the $W=0$ region. As the system size increases, the area of this complex loop expands. When the system size becomes sufficiently large, such as $N=150$ in Fig.~\ref{Fig2}(a), the complex loop extends beyond the inner loop of the PBC spectrum and encroaches into the $W=-1$ region, whereas the previously real spectrum remains real.
	
	This size-dependent spectral transition of the complex energy loop under OBCs is also accompanied by a transition in the skin localization of eigenstates. To characterize this transition, we calculate the spatial profiles $|\psi_{j,\alpha}^{(E)}|$ on the $A$- and $B$ chains for all wave functions with eigenvalues $E$ from the complex energy loop under OBCs, as shown in Figs.~\ref{Fig2}(b) and \ref{Fig2}(c) for $N=30$ and $N=150$, respectively. Here, $\psi_{j,\alpha}^{(E)}$ denotes the amplitude of the eigenstate with eigenvalue $E$ at the $j$th unit cell on sublattice $\alpha=\{A,B\}$. For small system sizes, each wave function with an eigenenergy $E$ in the $W=0$ region exhibits the CBSE, namely, it is simultaneously localized at opposite boundaries of the two chains [see Fig.~\ref{Fig2}(b)]. However, the CBSE is found to be unstable as the system size increases: the complex energy loop under OBCs eventually migrates into the $W=-1$ region, with the CBSE gradually vanishing and turning into unipolar NHSE localized at the right boundary for both chains [see Fig.~\ref{Fig2}(c) for $N=150$]. We note a fundamental distinction between the skin states observed in our model and the conventional critical skin effect found in typical non-Hermitian systems. In standard critical NHSE systems, eigenstates display the hallmark scale-free skin effect, characterized by a universal wavefunction profile and a localization length that diverges with increasing system size—leading to the collapse of rescaled wavefunctions onto a single curve and the eventual disappearance of the skin effect in the thermodynamic limit \cite{li2020nat,guo2021exact,li2021impurity}. In contrast, our system exhibits two distinct types of skin states: concurrent bipolar skin states at small system sizes and unipolar skin states at large system sizes. Both types show a clear breakdown of scale-free behavior. Specifically, the localization profiles in our model do not collapse onto a universal curve when rescaled by system size, and the inverse localization lengths remain finite or exhibit non-universal scaling (see Appendix B for details).
		 	
	The spectral features and NHSE for large $N$ can be further clarified by employing non-Bloch band theory within the framework of the generalized Brillouin zone~\cite{Yokomizo2019L}, which provides an analytical solution of non-Hermitian systems in the thermodynamic limit under OBCs. By substituting $ e^{ik}\to \beta$, the momentum-space Hamiltonian matrix becomes  
	\begin{equation}
		h(\beta) = \begin{pmatrix}
			J_1\beta + J_2\beta^{-1} & M \\
			M & J_3\beta + J_4\beta^{-1}
		\end{pmatrix}. \label{eq3}
	\end{equation}
	For a given energy $E$, the characteristic function $f(\beta, E) \equiv \det[h(\beta) - E]=0$ yields four solutions for $\beta$: $\beta_1$, $\beta_2$, $\beta_3$, and $\beta_4$. The generalized Brillouin zone is defined by the condition $|\beta_2| = |\beta_{3}|$ with the ordering $|\beta_1|\le|\beta_2|\le|\beta_3|\le|\beta_4|$. The corresponding energies $E$ that satisfy this condition constitute the OBC spectrum in the thermodynamic limit, as shown by the purple dots in Fig. \ref{Fig2}(a). It can be seen that in the thermodynamic limit, the complex energy loop under OBCs extends beyond the inner loop of the PBC spectrum and resides in the $W=-1$ region. This indicates that as the system size approaches the thermodynamic limit, all wave functions exhibit a stable right-directed NHSE.
	
	To analyze the CBSE in a small system, we define the normalized sublattice-dependent density imbalance for a given energy $E$ as
	\begin{align}
		I_{\alpha}^{(E)} = \frac{\sum_{j = 1}^{N} \mathrm{sgn}(j - N/2) |\psi_{j,\alpha}^{(E)}|^2}{\sum_{j = 1}^{N}|\psi_{j,\alpha}^{(E)}|^2}, \label{eq4}
	\end{align}
	where the total number of unit cells $N$ is chosen to be even. The sum and product of $I_{\alpha}^{(E)}$ for the two sublattices are defined as  $I_{S}^{(E)}=\sum_{\alpha}I_{\alpha}^{(E)}$ and $I_{P}^{(E)}= \prod_{\alpha}I_{\alpha}^{(E)}$, respectively. For extended states, both $I_{S}^{(E)}$ and $I_{P}^{(E)}$ approach zero. For CBSE, $I_{S}^{(E)}\approx 0$ while $I_{P}^{(E)}<0$ remains finite. In contrast, for unipolar NHSE, both $I_{S}^{(E)}$ and $I_{P}^{(E)}>0$ take finite values. Specifically, a negative (positive) value of $I_{S}^{(E)}$ indicates a left (right)-localized skin state. Figures \ref{Fig2}(d) and \ref{Fig2}(e) show the defined quantities for different eigenvalues with $N=30$ and $N=150$, respectively. For $N=30$, both $I_{S}^{(E)}$ and $I_{P}^{(E)}$ are positive for relatively large $|\mathrm{Re}(E)|$, indicating that the corresponding states are right-localized. 
	In contrast, states in the central region of $\mathrm{Re}(E)$, where $I_{S}^{(E)}\approx 0$ and $I_{P}^{(E)}<0$, exhibit bipolar skin characteristics. For the larger system size $N=150$ shown in Fig. \ref{Fig2}(e), both $I_{S}^{(E)}$ and $I_{P}^{(E)}$ are finite and positive for all states, indicating that	all states become right-directed skin states. 
	
	To further illustrate the size-dependent behavior, we plot $I_{S}^{(E)}$ and $I_{P}^{(E)}$ as functions of $\mathrm{Re}(E)$ and $N$ in Figs. \ref{Fig3}(a) and \ref{Fig3}(b), respectively. For small system sizes, $I_{S}^{(E)} \approx 0$ and $I_{P}^{(E)}<0$ in the central region of $\mathrm{Re}(E)$, corresponding to the emergence of CBSE, while $I_{S}^{(E)}$ and $I_{P}^{(E)}$ are positive when $|\mathrm{Re}(E)|\gtrsim 1.63$, indicating the presence of right-localized skin states. As the system size $N$ increases, the central region featuring CBSE gradually shrinks, and once $N$ exceeds a critical value (approximately at $N=86$), all the states become right-localized. 
	
	Furthermore, we compute the winding number for each eigenstate, $W_{\mathrm{OBC}}^{(E)}$, by taking its corresponding eigenvalue under OBCs as the reference energy in Eq. (\ref{eq2}). As demonstrated in Refs. \cite{Yokomizo2019L,Zhang2020}, there exits a direct correspondence between $W_{\mathrm{OBC}}^{(E)}$ and the emergence of unipolar skin modes in non-Hermitian systems. Specifically, when the reference energy $E_0$ lies within a region where the winding number, as determined by Eq. \eqref{eq2}, is nonzero, the associated mode with eigenvalue $E_0$ exhibits a unipolar NHSE. In contrast, a vanishing winding number, $W_{\mathrm{OBC}}^{(E)}=0$, signifies the absence of the unipolar NHSE. Figure \ref{Fig3}(c) presents the values of $W_{\mathrm{OBC}}^{(E)}$ for various eigenstates and different system sizes under OBCs. We observe that $W_{\mathrm{OBC}}^{(E)}=-1$ is always satisfied when $|\mathrm{Re}(E)|\gtrsim 1.63$, indicating that all wave functions in these regions exhibit right-directed skin characteristics. In contrast, wave functions near the center of $\mathrm{Re}(E)$ display bipolar characteristics with $W_{\mathrm{OBC}}^{(E)}=0$. As the system size increases, the center region of $\mathrm{Re}(E)$ where $W_{\mathrm{OBC}}^{(E)}=0$ gradually shrinks, and eventually vanishes when $N$ exceeds approximately $86$. Finally, we count the number $N_{W}$ of distinct winding numbers $W_{\mathrm{OBC}}^{(E)}$ for all OBC eigenenergies $E$ at a given system size $N$, and display the results in Fig. \ref{Fig3}(d). It is seen that $N_{W}=2$ for small system sizes, indicating the coexistence of CBSE and right-directed NHSE that possess two different values of $W_{\mathrm{OBC}}^{(E)}$. When $N>86$, a size-dependent skin transition occurs, after which $N_W=1$ and all eigenstates become unipolar NHSE states.

	\subsection{Tangent loops with $\delta_{b}=\delta_{a}$}
	
	\begin{figure}[htbp]
		\includegraphics[width=0.5\textwidth]{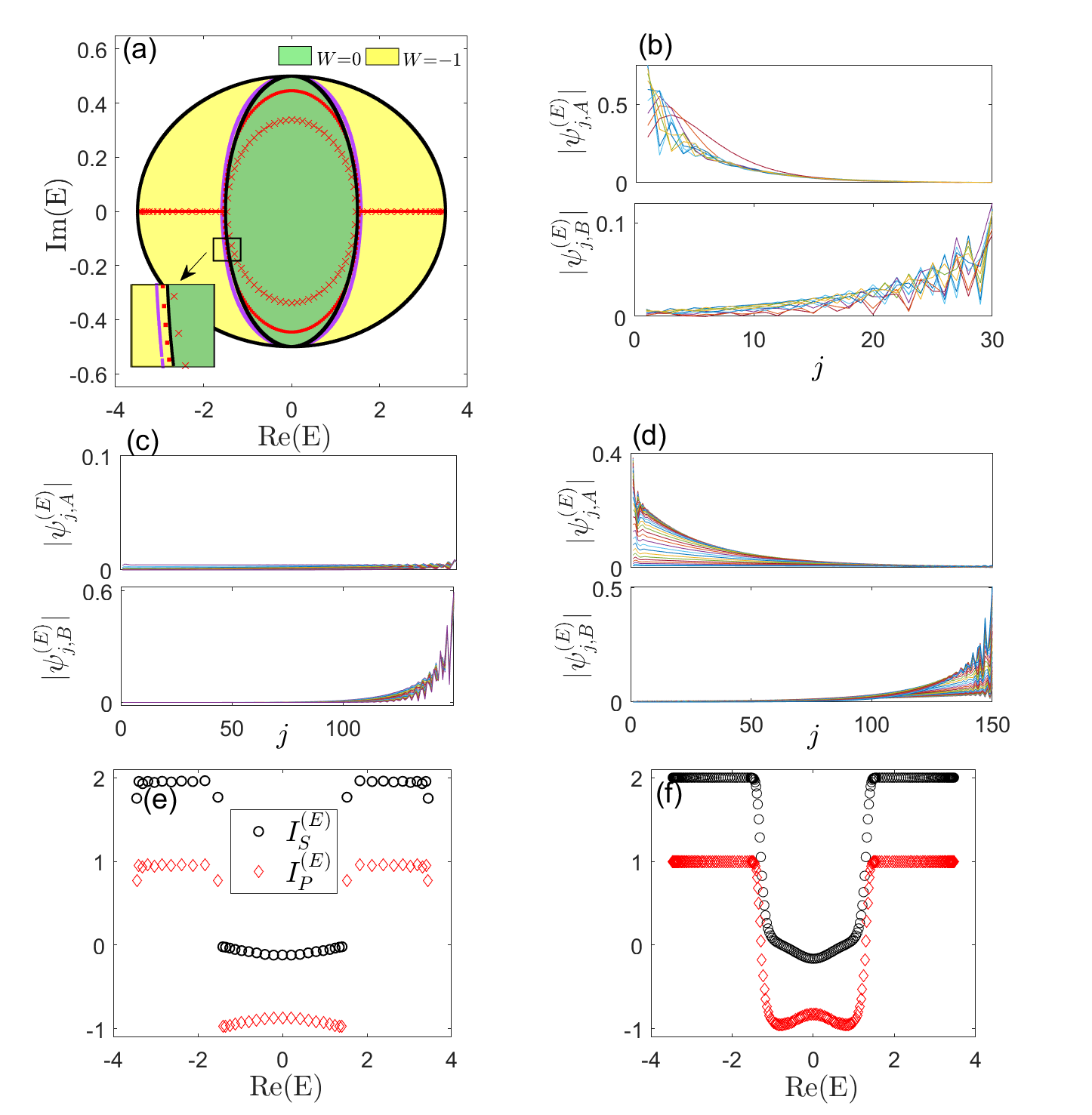}
		\caption{(a) Energy spectra for $\delta_{b}=\delta_{a}$ under PBCs and OBCs. Black curves represent the spectrum under PBCs, red crosses denote the OBC spectrum for $N = 30$, red dots correspond to $N = 150$, and purple dots indicate the thermodynamic-limit spectrum obtained from the non-Bloch band theory. The yellow (green) shading marks the $W = -1$ ($W = 0$) region. (b) The spatial profiles $|\psi_{j,A}^{(E)}|$ and $|\psi_{j,B}^{(E)}|$ of all wave functions with eigenvalue $E$ from the complex energy loop under OBCs for $N = 30$. (c) and (d) The spatial profiles $|\psi_{j,A}^{(E)}|$ and $|\psi_{j,B}^{(E)}|$ of all wave functions under OBCs for $N = 150$ with eigenvalue $E$ resided in the $W=-1$ and $W=0$ regions, respectively. The quantities $I_S^{(E)}$ and $I_P^{(E)}$ for different eigenvalues for (e) $N = 30$ and (f) $N = 150$. Here, $J_1 = 0.5$, $J_3 = 2$, $\delta_a = 0.5$, $\delta_b = 0.5$, and $M = 0.01$.}
		\label{Fig4}
	\end{figure} 

	\begin{figure}[htbp]
		\includegraphics[width=0.5\textwidth]{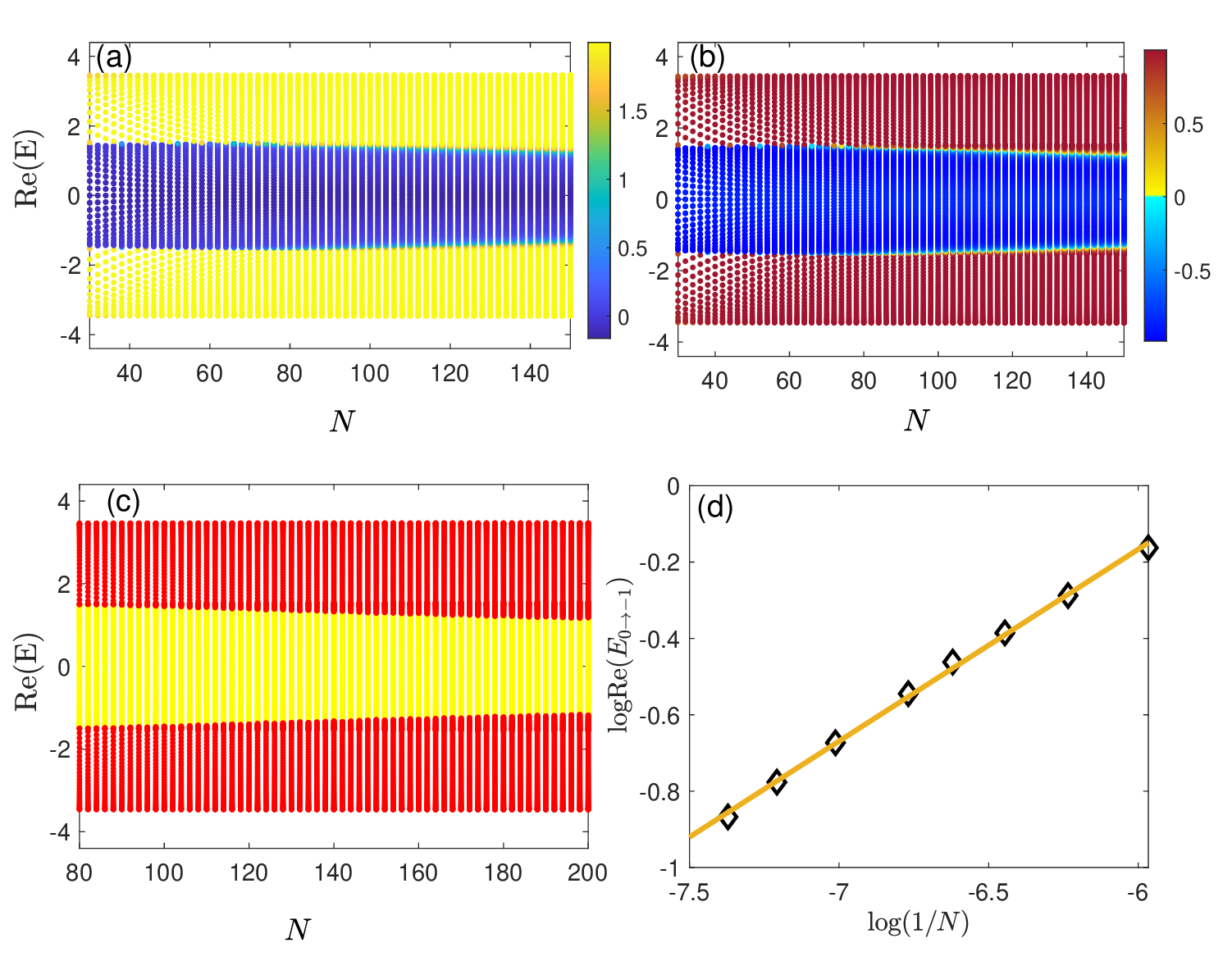}
		\caption{(a) $I_S^{(E)}$ and (b) $I_P^{(E)}$ as functions of $\mathrm{Re}(E)$ and system size $N$. (c) The values of $W_{\mathrm{OBC}}^{(E)}$ for OBC eigenstates and different system size $N$. Yellow dots mark the eigenenergies with $W_{\mathrm{OBC}}^{(E)}=0$, and red dots indicates the eigenenergies with $W_{\mathrm{OBC}}^{(E)}=-1$. (d) The fitting of the transitions, where $W_{\mathrm{OBC}}^{(E)}$ changes from zero to nonzero, as a function of $1/N$. Here, $J_1 = 0.5$, $J_3 = 2$, $\delta_a = 0.5$, $\delta_b = 0.5$, and $M = 0.01$.}
		\label{Fig5}
	\end{figure}
	
	Next, we consider the critical parameter regime where $\delta_a=\delta_b$. In this case, the PBC energy spectrum forms two complex loops [black curves in Fig.~\ref{Fig4}(a)] that touch at two energy points $\tilde{E}^{\prime}_1=-i\sqrt{\delta_a^2-M^2}$ and $\tilde{E}^{\prime}_2=i\sqrt{\delta_a^2-M^2}$ in the complex energy plane. Overall, the system at this critical point behaves similarly to the case with $\delta_b>\delta_a$; the PBC spectrum consists of two loops and supports only the spectral winding numbers $W=0$ and $W=-1$, although the two loops now touch at the aforementioned points. Under OBCs, the spectrum features a central loop exhibit CBSE in a small size system [e.g., $N=30$, see Fig.~\ref{Fig4}(b)]. In the thermodynamic limit, however, the OBC spectrum for $\delta_a=\delta_b$ [purple dots in Fig.~\ref{Fig4}(a)] predominantly lies in the region with $W=-1$, except at the two touching points $\tilde{E}^{\prime}_1$ and $\tilde{E}^{\prime}_2$, indicating the domination of right-directed NHSE.
	
	Nevertheless, when $\delta_b=\delta_a$, the system shows markedly different asymptotic behavior as it approaches the thermodynamic limit. As shown in Fig. \ref{Fig3}(a), the central loop of the OBC spectrum at this critical point remains in the region with $W=0$ even when the system size increases to $N=150$; in contrast, for $\delta_b>\delta_a$, the CBSE has already vanished at this scale (see Fig. \ref{Fig2}). Consistently, the corresponding eigenstates exhibit the CBSE, as verified by their spatial distributions in Fig. \ref{Fig4}(d). Meanwhile, other eigenstates with complex eigenvalues outside the inner PBC loop exhibit right-directed NHSE, as illustrated in Fig. \ref{Fig4}(c). Furthermore, in Figs. \ref{Fig4}(e) and (f), we plot the sum and product of density imbalance, $I_{S}^{(E)}$ and $I_{P}^{(E)}$, for all eigenstates at $N=30$ and $N=150$, respectively, both of which qualitatively display the same behavior.
	
	To further demonstrate the asymptotic behavior, Fig. \ref{Fig5} shows $I_{S}^{(E)}$, $I_{P}^{(E)}$, and the values of $W_{\rm{OBC}}^{(E)}$ for all OBC eigenstates as functions of the system size $N$. For small $N$, the system exhibits clear signatures of CBSE, i.e., $I_S^{(E)}\to 0$, $I_P^{(E)}<0$, and $W_{\rm{OBC}}^{(E)}=0$ in the region of $|{\rm Re}[E]|<1.53$. As $N$ increases, the CBSE region gradually shrinks, albeit much more slowly compared to the case with $\delta_b>\delta_a$ [see Fig. \ref{Fig3}(c)], and eventually tends to vanish in the thermodynamic limit. As shown in Fig. \ref{Fig5}(d), we fit the transition points, where $W_{\rm{OBC}}^{(E)}$ changes from zero to nonzero, as a function of $1/N$. The fitting reveals a power-law decay as the system size increases. In the limit of $N\to\infty$, the transition approaches ${\rm Re}(E_{-1\leftrightarrow0})\to 0$.

	\subsection{Intersecting loops with $\delta_{b}<\delta_{a}$}
	
	\begin{figure*}[htbp]
		\includegraphics[width=1\textwidth]{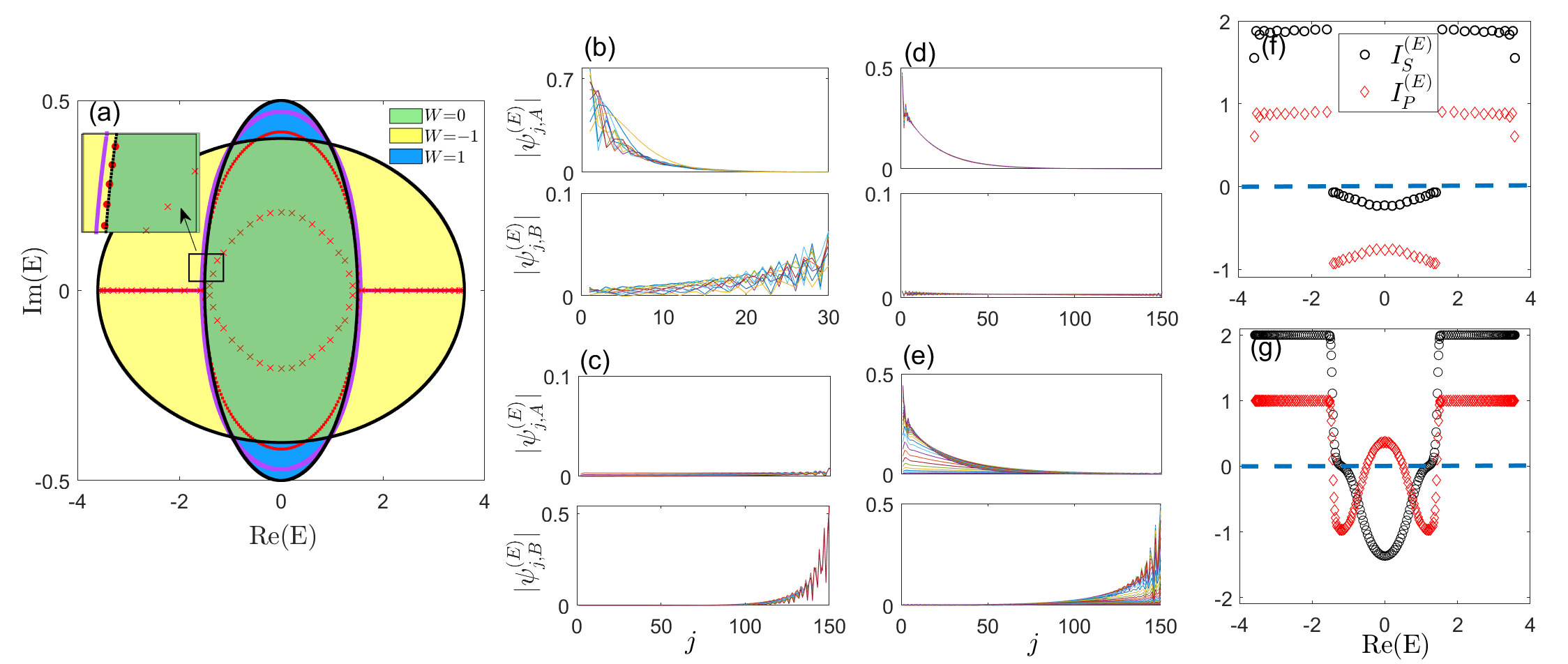}
		\caption{(a) Energy spectra for $\delta_{b}<\delta_{a}$ under PBCs and OBCs. Black curves represent the spectrum under PBCs, red crosses denote the OBC spectrum for $N = 30$, red dots correspond to $N = 150$, and purple dots indicate the thermodynamic-limit spectrum obtained from the non-Bloch band theory. The yellow, green, and blue shadings mark the $W = -1$, $0$, and $1$ regions, respectively. (b) The spatial profiles $|\psi_{j,A}^{(E)}|$ and $|\psi_{j,B}^{(E)}|$ of all wave functions with eigenvalue $E$ from the complex energy loop under OBCs for $N = 30$. (c)-(e) The spatial profiles $|\psi_{j,A}^{(E)}|$ and $|\psi_{j,B}^{(E)}|$ of all wave functions under OBCs for $N = 150$ with eigenvalue $E$ resided in the $W=-1$, $W=1$, and $W=0$ regions, respectively. The quantities $I_S^{(E)}$ and $I_P^{(E)}$ for different eigenvalues for (f) $N = 30$ and (g) $N = 150$. Here, $J_1 = 0.5$, $J_3 = 2$, $\delta_a = 0.5$, $\delta_b = 0.4$, and $M = 0.01$.}
		\label{Fig6}
	\end{figure*}  
	
	\begin{figure}[htbp]
		\includegraphics[width=0.5\textwidth]{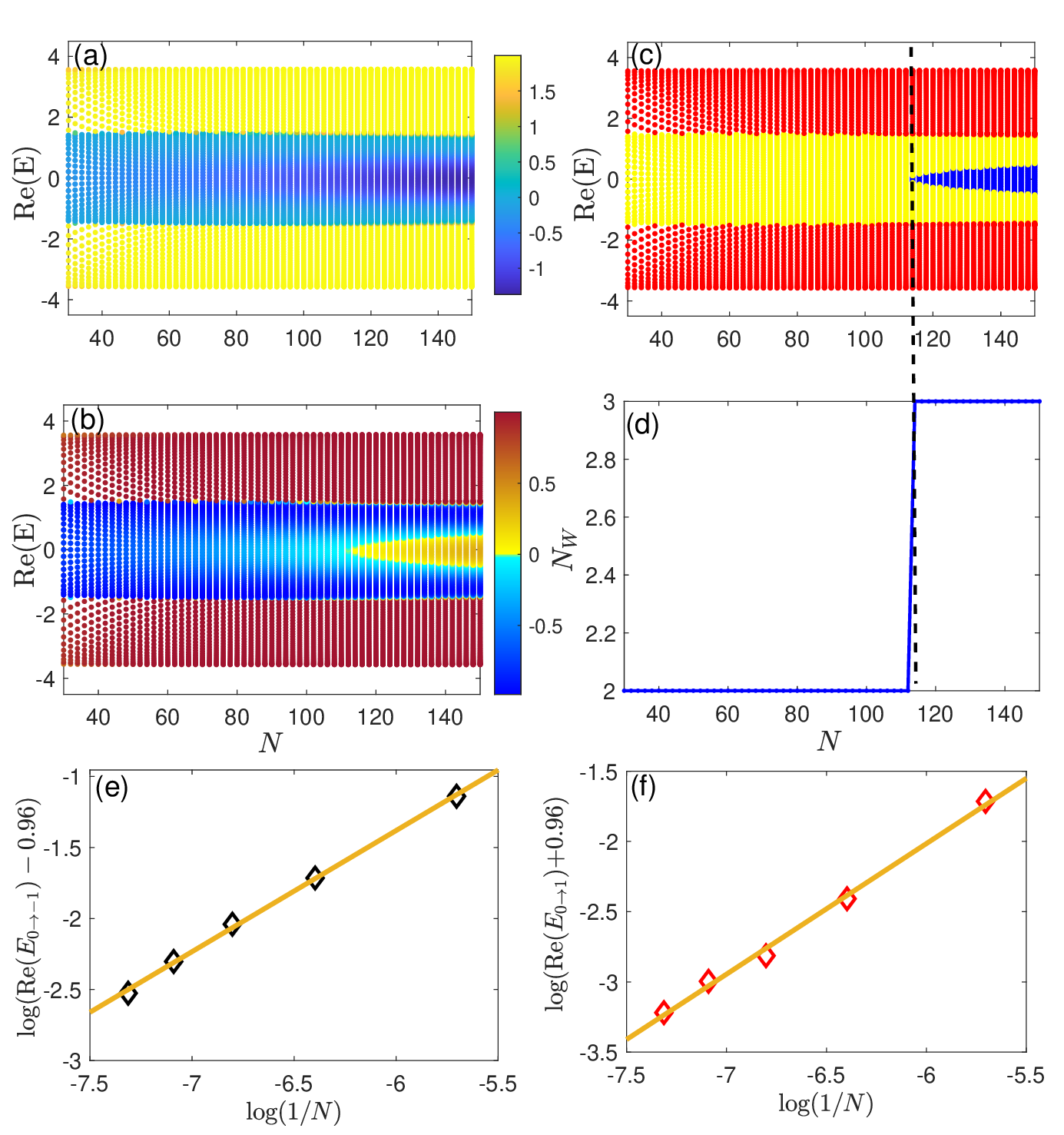}
		\caption{(a) $I_S^{(E)}$ and (b) $I_P^{(E)}$ as functions of $\mathrm{Re}(E)$ and system size $N$. (c) The values of $W_{\mathrm{OBC}}^{(E)}$ for various eigenstates and different system size $N$ under OBCs. The yellow dot marks the eigenstate with $W_{\mathrm{OBC}}^{(E)}=0$, the red dot indicates the eigenstate with $W_{\mathrm{OBC}}^{(E)}=-1$, and the blue dot represents the eigenstate with $W_{\mathrm{OBC}}^{(E)}=1$. (d) The number $N_W$ of distinct winding numbers $W_{\mathrm{OBC}}^{(E)}$ as a function of system size $N$. The dashed lines in both (c) and (d) correspond to $N = 114$. (e) The fitting of the size-dependent transitions where $W_{\mathrm{OBC}}^{(E)}$ changes from $0$ to $-1$. (f) The fitting of the size-dependent transitions where $W_{\mathrm{OBC}}^{(E)}$ changes from $0$ to $1$. Here, $J_1 = 0.5$, $J_3 = 2$, $\delta_a = 0.5$, $\delta_b = 0.4$, and $M = 0.01$.}
		\label{Fig7}
	\end{figure}  
	
	When $\delta_a>\delta_b$, the previously inner energy loop under PBCs expands and extends beyond the outer loop, resulting in four intersection points. As shown in Fig.~\ref{Fig6}(a), the complex energy plane is thus divided into three distinct regions, each characterized by a different winding number: $W=-1$, $0$, and $1$, respectively. 
		
	Similar to previous cases, the OBC spectrum consists of real eigenenergies at relatively large $|{\rm Re}(E)|$, where eigenstates display a size-independent, right-directed NHSE. In the central region, the spectrum features complex eigenenergies that form a loop and undergo a size-dependent skin transition. Specifically, for small system sizes, the complex OBC eigenenergies are entirely contained within the region enclosed by both PBC energy loops and exhibits the CBSE, as illustrated by the profiles $|\psi_{j,\alpha}|$ of all states with complex eigenvalues in Fig.~\ref{Fig6}(b) for $N=30$. As the system size increases, the OBC loop expands and partially enters regions with $W=1$ and $W=-1$ at intermediate sizes, as shown in Fig.~\ref{Fig6}(a) for $N=150$. Figures \ref{Fig6}(c)-6(e) show the profiles $|\psi_{j,\alpha}|$ of all states of the complex OBC spectrum residing in the $W=-1$, $W=1$, and $W=0$ regions, respectively, for $N=150$. The states in the $W=1$ ($W=-1$) regions exhibit left- (right-) directed localization, while those in the $W=0$ region continue to display CBSE. In the thermodynamics limit, as indicated by the OBC spectrum obtained from non-Bloch band theory [purple dots in Fig. \ref{Fig6}(a)], all eigenstates fall in the regions with either $W=1$ or $W=-1$, manifesting the conventional bipolar NHSE.
	
	Figures \ref{Fig6}(f) and \ref{Fig6}(g) display $I_{S}^{(E)}$ and $I_{P}^{(E)}$ as functions of $\mathrm{Re}(E)$ for $N=30$ and $N=150$, respectively. For small system sizes ($N=30$), both $I_{S}^{(E)}$ and $I_{P}^{(E)}$ are finite and positive for $|\mathrm{Re}(E)|\gtrsim1.57$, indicating that the corresponding wave functions are right localized. In the central region of $\mathrm{Re}(E)$, $I_{S}^{(E)}\approx 0$ and $I_{P}^{(E)}<0$, signaling the presence of CBSE. For an intermediate system size, as shown in Fig. \ref{Fig6}(g) for $N=150$, the wave functions for $|\mathrm{Re}(E)|\gtrsim1.43$ maintain their right-localized character. In contrast, the region with $|\mathrm{Re}(E)|\lesssim1.43$ is further subdivided:
	when $|\mathrm{Re}(E)|\lesssim0.48$, we find $I_{S}^{(E)}<0$ and $I_{P}^{(E)}>0$, indicating left-localized wave functions; for $0.48<|\mathrm{Re}(E)|\lesssim1.43$, $I_{S}^{(E)}\approx 0$ and $I_{P}^{(E)}<0$, indicating the persistence of CBSE. These results demonstrate that for intermediate system sizes, the system exhibits the coexistence of conventional bipolar NHSE and CBSE. To further characterize the size-dependent behavior, we present $I_{S}^{(E)}$ and $I_{P}^{(E)}$ as functions of $\mathrm{Re}(E)$ and $N$ in Figs. \ref{Fig7}(a) and \ref{Fig7}(b), respectively. For small $N$, the system displays the coexistence of CBSE and right-directed NHSE, as diagnosed by the behaviors of $I_{S}^{(E)}$ and $I_{P}^{(E)}$. As $N$ increases, the central region exhibiting CBSE gradually shrinks. Once $N$ exceeds a certain threshold, a region in the central part of $\mathrm{Re}(E)$ with $I_{S}^{(E)}<0$ and $I_{P}^{(E)}>0$ emerges, corresponding to the appearance of left-directed states that coexist with the right-directed and concurrent bipolar skin states. Figures ~\ref{Fig7}(c) and \ref{Fig7}(d) illustrate the winding numbers of different OBC eigenstates and the number $N_W$ of distinct winding numbers $W_{\mathrm{OBC}}^{(E)}$ among all OBC eigenstates, respectively, as functions of the system size $N$. In contrast to previous cases, the system now support a third region with $W=1$ and $N_W=3$ when $N\gtrsim 114$, corresponding to the coexistence of CBSE and conventional bipolar NHSE (i.e., left- and right-directed NHSEs). Additionally, the region with $W_{\mathrm{OBC}}^{(E)}=0$ asymptotically disappears only when $N\rightarrow \infty$, similar to the critical scenario with $\delta_a=\delta_b$. As seen in Figs. \ref{Fig7}(e) and \ref{Fig7}(f), we fit the size-dependent transitions where $W_{\mathrm{OBC}}^{(E)}$ changes from $0$ to $-1$ and from $0$ to $1$ as functions of system size $1/N$, respectively. Both fits exhibit power-law decay behavior as the system size increases. In the limit $N\to\infty$, the region with $W_{\mathrm{OBC}}^{(E)}=0$ vanishes, and the transitions from $W_{\mathrm{OBC}}^{(E)}=-1$ to $W_{\mathrm{OBC}}^{(E)}=1$ occur at $\mathrm{Re}(E_{-1\leftrightarrow1})\approx \pm 0.96$.
	
\begin{figure}[t]
		\includegraphics[width=0.5\textwidth]{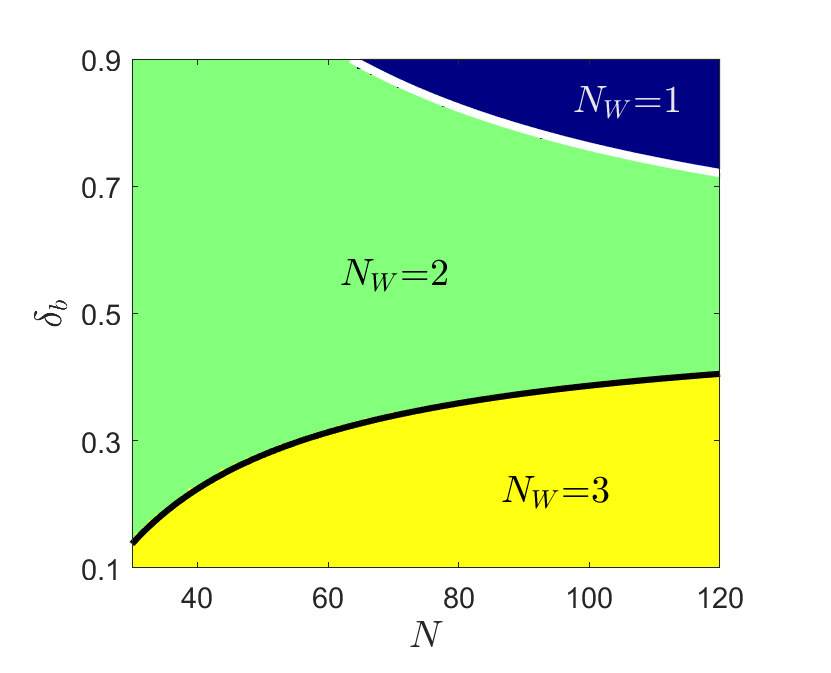}
		\caption{Transition diagram of $N_W$ as a function of $\delta_b$ and system size $N$, with $J_1=0.5$, $J_3=2$, $\delta_a=0.5$, and $M=0.01$. The yellow region corresponds to $N_W = 3$, the blue region to $N_W = 1$, and the green region to $N_W = 2$.}
		\label{Fig8}
	\end{figure}

	To provide a comprehensive view of the size-dependent skin transition, we present a transition diagram of the number of distinct winding numbers $N_W$ as functions of the system size $N$ and $\delta_b$ in Fig. \ref{Fig8}. In the region $\delta_a<\delta_b<1$, the system undergoes a size-dependent skin transition from the coexistence of CBSE and right-directed NHSE (green region with $N_W=2$) to a purely unipolar NHSE regime (blue region with $N_W=1$) as the system size increases. The corresponding transition line can be well fitted by $\delta_b = 27.79/(N+5.921)+0.5$, as indicated by the white line. Another transition line, fitted by $\delta_b=-11.56/(N+1.917)+0.5$ (black line), describes the transition from $N_{W}=2$ to $N_{W}=3$.	
	
	\section{Skin transitions for finite interchain couplings}
	
		\begin{figure}[htbp]
		\includegraphics[width=0.5\textwidth]{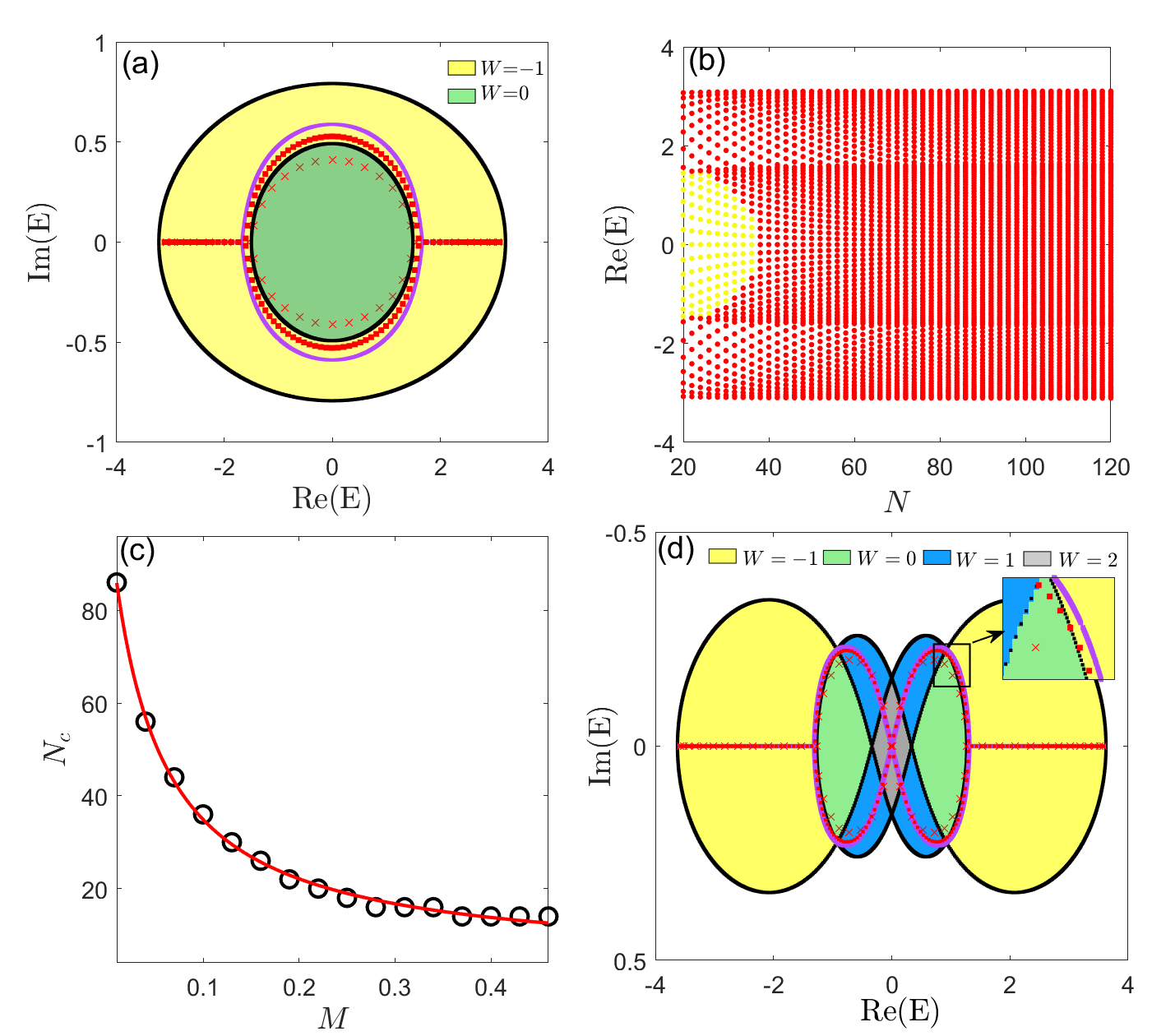}
		\caption{(a) Energy spectra of the system under PBCs and OBCs with $\delta_b=0.8,M=0.1$. Black curves represent the PBC spectrum, red crosses denote OBC spectra for $N = 20$, red dots for $N = 60$, and purple dots indicate the thermodynamic-limit spectrum obtained from the non-Bloch band theory. The yellow and green shaded designate the $W = -1$ and $0$ region, respectively. (b) The values of $W_{\mathrm{OBC}}^{(E)}$ for various eigenstates and different system size $N$ with $\delta_b=0.8,M=0.1$ under OBCs. The yellow dot marks the eigenstate with $W_{\mathrm{OBC}}^{(E)}=0$, and the red dot indicates the eigenstate with $W_{\mathrm{OBC}}^{(E)}=-1$. (c) The critical system size of the size-dependent skin transition $N_c$ as a function of $M$ with $\delta_b=0.8$. (d) Energy spectra of the system under PBCs and OBCs with $\delta_b=0.6$ and $M=0.7$. Black curves represent the PBC spectrum, red crosses denote OBC spectra for $N = 30$, red dots for $N = 120$, and purple dots indicate the thermodynamic-limit spectrum obtained from the non-Bloch band theory. The yellow, green, blue, and gray shaded regions represent the $W = -1$, $0$, $1$, and $2$, respectively. Here, $J_1=0.5$, $J_3=2$, and $\delta_a=0.5$.}\label{Fig9}
	\end{figure} 
	
	So far, we have systematically revealed the size-dependent skin transition in the weak coupling limit with $M=0.01$. In this regime, CBSE emerges in the region with $W=0$ for small system sizes, and gradually evolves into left- or right-directed NHSEs with $W\neq0$ as the system size increases.	In this section, we take $\delta_b>\delta_a$ as an example to discuss the instability of the $W=0$ region under stronger interchain coupling, where the coupling strength becomes comparable to other parameters.
	
	Figure \ref{Fig9}(a) presents the energy spectra of the system with $M=0.1$. The black curves represent the spectrum under PBCs, while the red crosses and red dots correspond to the OBC spectra for $N=20$ and $N=60$, respectively. Both the PBC and OBC spectra exhibit features similar to those observed for $M=0.01$. Specifically, the PBC spectrum consists of two loops, resulting in distinct regions with $W=0$ and $-1$ in the complex energy plane. As shown in Fig. \ref{Fig9}(a), for $N=20$ (red crosses), all eigenvalues in the $W=-1$ region are real, while the remaining eigenvalues, forming a complex loop, are located within the $W=0$ region. When $N=60$, the previously complex OBC spectrum (red dots) extends into the $W=-1$ region. In the thermodynamic limit, the OBC spectrum obtained from non-Bloch band theory, shown as purple dots in Fig. \ref{Fig9}(a), lies entirely within the $W=-1$ region, indicating the emergence of right-directed NHSE. To further characterize this size-dependent transition, we plot $W_{\mathrm{OBC}}^{(E)}$ as functions of the real part of the energy spectrum $\mathrm{Re}(E)$ and the system size $N$ in Fig. \ref{Fig9}(b). Similar to the case of $M=0.01$, for $M=0.1$ and small system sizes, the states at the edges of $\mathrm{Re}(E)$ correspond to $W_{\mathrm{OBC}}^{(E)}=-1$, while those in the central region  have $W_{\mathrm{OBC}}^{(E)}=0$. As the system size increases, the central region with $W_{\mathrm{OBC}}^{(E)}=0$ gradually shrinks and disappears when $N\gtrsim36$. Figure \ref{Fig9}(c) shows the critical system size $N_c$ for this transition as a function of the interchain coupling amplitude $M$. It can be seen that as $M$ increases, the critical size $N_c$ decreases monotonically, indicating that the instability of the $W=0$ region is enhanced with increasing $M$, as long as the nested spectral structure remains unchanged. As the inter-chain coupling $M$ increases further, the spectral structure undergoes a significant transformation (see Appendix C). Notably, the critical system size $N_c$ displays a discontinuous jump, reflecting the abrupt change in the spectral topology. In Fig. \ref{Fig9}(d), we present an example with $M=0.7$, where the PBC spectrum forms two figure-eight loops that partially overlap, resulting in four topologically distinct regions with $W=0,\pm1,$ and $2$, respectively. Nevertheless, we find that the system supports OBC eigenenergies in the $W=0$ region only when the system size is small (e.g., $N=30$ in the figure). As the system size increases to $N=120$, all eigenstates are located in regions with $W\neq0$, indicating the presence of unipolar NHSEs for their corresponding eigenstates. This observation demonstrates the universality of the size-dependent transition in our system, which persists beyond the weak-coupling regime. 
	
	\section{Experimental Proposal}
	
	\begin{figure}[htbp]
		\includegraphics[width=0.5\textwidth]{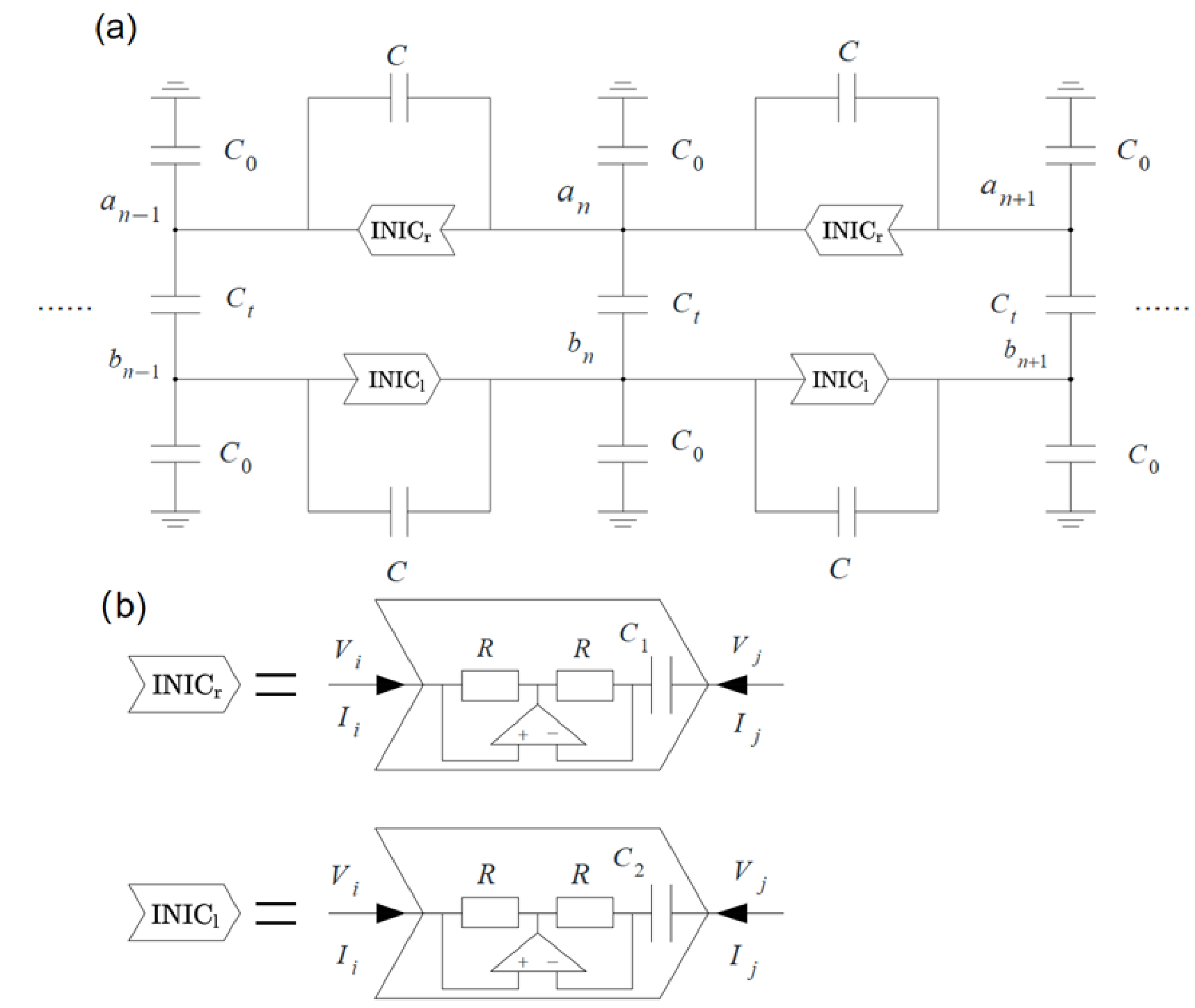}
		\caption{(a)  Electrical circuit implementation of the model described in Eq.~\eqref{eq1}. (b) Detailed schematic of $\mathrm{INIC}_r$ and $\mathrm{INIC}_l$.}
		\label{Fig13}
	\end{figure} 
	
	These two weakly coupled nonreciprocal chains can be experimentally realized using electrical circuits. We designed a non-Hermitian electrical circuit, as illustrated in Fig. \ref{Fig13}, which corresponds to the model presented in Eq. \eqref{eq1}. In Fig. \ref{Fig13}, nonreciprocal hopping is achieved through negative impedance converters (INICs) that invert current, as shown in Fig. \ref{Fig13}(b). Specifically, the nonreciprocal hopping between $a_n$ ($b_n$) and $a_{n+1}$ ($b_{n+1}$) is simulated by combining a conventional capacitor $C$ with an $\mathrm{INIC}_r$ for rightward hopping or $\mathrm{INIC}_l$ for leftward hopping. The INIC consists of a capacitor $C_1$ ($C_2$), an operational amplifier, and two resistors of equal resistance $R$. When current flows from left to right (or right to left), the effective capacitance between $a_n$ ($b_n$) and $a_{n+1}$ ($b_{n+1}$) becomes $C+C_1$ ($C-C_2$), respectively. The on-site potential at each site is simulated by grounding a capacitor $C_0$. The inter-chain coupling amplitude can be tuned by using a coupling capacitor $C_t$. The circuit Laplacian is defined as the response of the grounded-voltage vector $\mathbf{V}$ to the input current vector $\mathbf{I}$ by
		\begin{equation}
			\mathbf{I}(\omega)=\mathcal{J}(\omega)\,\mathbf{V}(\omega).
			\label{eq:laplacian_def}
		\end{equation}
	Using Eq.~\eqref{eq:laplacian_def}, the current of each node within a unit cell can be expressed as
	\begin{align}
		I_{a_n}
		&= i\omega \Big[
		(C{+}C_1)\big(V_{n-1}^{a}-V_n^{a}\big)
		+ C_t\big(V_n^{b}-V_n^{a}\big)
		\notag\\[-2pt]&\qquad
		+ (C{-}C_1)\big(V_{n+1}^{a}-V_n^{a}\big)
		+ C_0\,(0-V_n^{a})
		\Big], \\
		I_{b_n}
		&= i\omega \Big[
		(C{-}C_2)\big(V_{n-1}^{b}-V_n^{b}\big)
		+ C_t\big(V_n^{a}-V_n^{b}\big)
		\notag\\[-2pt]&\qquad
		+ (C{+}C_2)\big(V_{n+1}^{b}-V_n^{b}\big)
		+ C_0\,(0-V_n^{b})
		\Big].
		\label{eq:jmatrix}
	\end{align}
	Consequently, the targeted circuit Laplacian $\mathcal{J}(\omega)$ simulates the model in Eq.~\eqref{eq:jmatrix} as follows:
\begin{equation}
	\label{eq:Jomega_topleft}
	\begingroup
	\scriptsize
	\setlength{\arraycolsep}{4pt}
	\renewcommand{\arraystretch}{1.0}
	\resizebox{\columnwidth}{!}{$%
		\mathcal{J}(\omega)=i\omega
		\left(
		\begin{array}{ccccc}
			0 & C_t         & C{-}C_1      & 0            & \cdots \\
			C_t         & 0 & 0            & C{+}C_2 & \cdots \\
			C{+}C_1   & 0              & 0 & C_t     & \cdots \\
			0              & C{-}C_2        & C_t       & 0 & \cdots \\
			\vdots         & \vdots         & \vdots       & \vdots       & \ddots
		\end{array}
		\right)-i\omega(2C+C_t+C_0)\mathbf{1}
		$}
	\endgroup
\end{equation}
	where $\mathbf{1}$ denotes the identity matrix. The energy spectrum of the system can be extracted from the admittance spectrum of the circuit, while the spatial distributions of states can be probed by measuring the voltage at each node.
	
	\section{Conclusion}
	
	In summary, we have systematically investigated size-dependent skin effect transitions in a system of two coupled nonreciprocal chains, while previous studies identified the critical NHSE in the weakly coupled regime, characterized by a size-dependent transition between scale-free localization and NHSE. By analyzing the size dependence of the complex energy spectra and  associated winding numbers under both PBCs and OBCs, we uncover diverse localization phenomena, including the coexistence and instability of CBSE in the $W=0$ region, as well as size-driven transitions between CBSE, unipolar NHSE, and conventional bipolar NHSE. Importantly, we find that the presence and size-dependent behavior of eigenstates in the $W=0$ region are universal features of certain non-Hermitian systems. These properties persist even as the interchain coupling increases beyond the weakly coupled regime. Our findings highlight the crucial role of system size and interchain coupling in determining the topological and localization properties of non-Hermitian systems, and provide valuable insights for the design and control of skin effects in engineered lattices.

	\section*{Appendix A: Parameter choices and generalization}
	
	\begin{figure}[htbp]
		\includegraphics[width=0.5\textwidth]{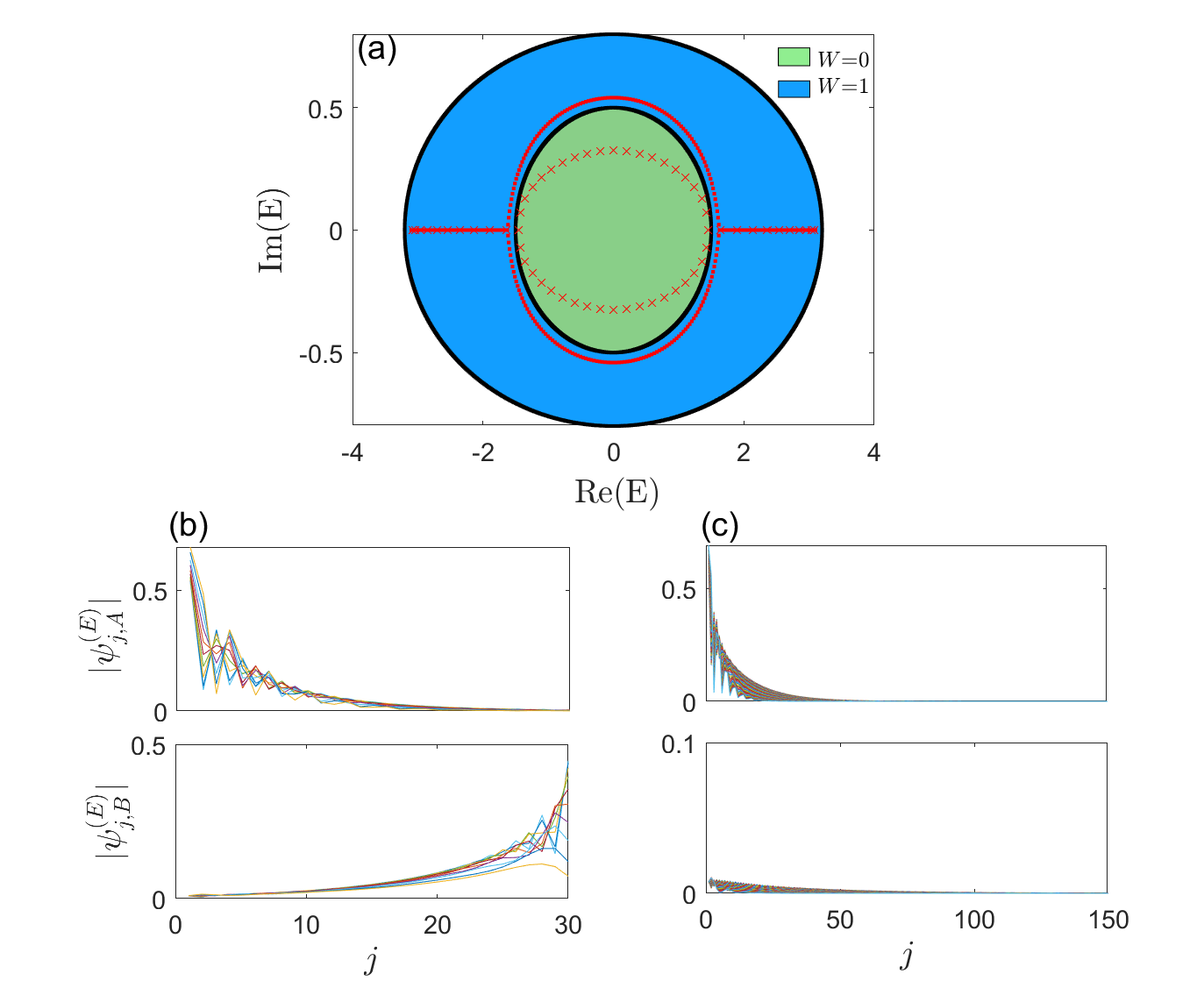}
		\caption{(a) Energy spectra for $\delta_{b}>\delta_{a}$ under PBCs and OBCs in the regime $J_3+J_4<J_1+J_2$. The black curves indicate the spectrum under PBCs, while red crosses and red dots represent the OBC spectra for $N=30$ and $N = 150$, respectively. (b), (c) Spatial profiles $|\psi_{j,A}^{(E)}|$ and $|\psi_{j,B}^{(E)}|$ of all wavefunctions corresponding to eigenvalue $E$ from the complex energy loop under OBCs for $N = 30$ and $N = 150$, respectively. Here, $J_1 = 1.2$, $J_3 = 1$, $\delta_a = 0.8$, $\delta_b = 0.5$, and $M=0.01$.}
		\label{Fig11}
	\end{figure} 
			
		In the main text, our analysis primarily focuses on the regime where $J_3+J_4>J_1+J_2$. To demonstrate the generality of our findings, we also consider the complementary case $J_3+J_4 < J_1+J_2$. This can be realized by swapping the hopping amplitudes between the $A$- and $B$ chains, i.e., $J_1\leftrightarrow J_4$ and $J_2 \leftrightarrow J_3$, which ensures $J_3+J_4 < J_1+J_2$. As shown in Fig. \ref{Fig11}(a), we plot the energy spectra under PBCs and OBCs with $J_1=1.2$, $J_2=2$, $J_3=1$, and $J_4=0.5$. In this case, the PBC spectrum consists of two nested loops---qualitatively similar to the scenario discussed in the main text. We compute the spectral winding numbers $W$ with respect to reference energies $E_0$ located within the PBC spectrum. The region enclosed by both loops (green shaded region) has $W=0$, whereas $E_0$ lying in the region enclosed by only the outer loop (blue shaded region) yields $W=1$. The OBC spectrum exhibits behavior analogous to that in the main text. For small system sizes---taking $N=30$ as an example---all energies in the $W=1$ region remains real, while the remaining states form a complex-energy loop confined to the $W=0$ region. As the system size increases, this complex loop expands. When $N$ exceeds a critical value $N_c=86$, the complex loop protrudes beyond the inner PBC loop and intrudes into the $W=1$ region, while the previously real spectrum remain real. Figures \ref{Fig11}(b) and \ref{Fig11}(c) display the spatial profiles $|\psi_{j,\alpha}^{(E)}|$ on the $A$- and $B$ chains for all eigenstates whose eigenvalues $E$ belong to the complex energy loop under OBCs. For $N=30$, each eigenstate with $E$ residing in the $W=0$ region exhibits a CBSE. However, when the system size exceeds $N_c$, this CBSE gradually disappears and evolves into a unipolar NHSE localized at the left boundary for both chains. Our calculations reveal that the CBSE is also unstable and undergoes a system-size dependent skin transition, leading to the same qualitative conclusions as in the main text.

	\section*{Appendix B: Scaling behavior of skin states}
	
	\begin{figure}[htbp]
		\includegraphics[width=0.5\textwidth]{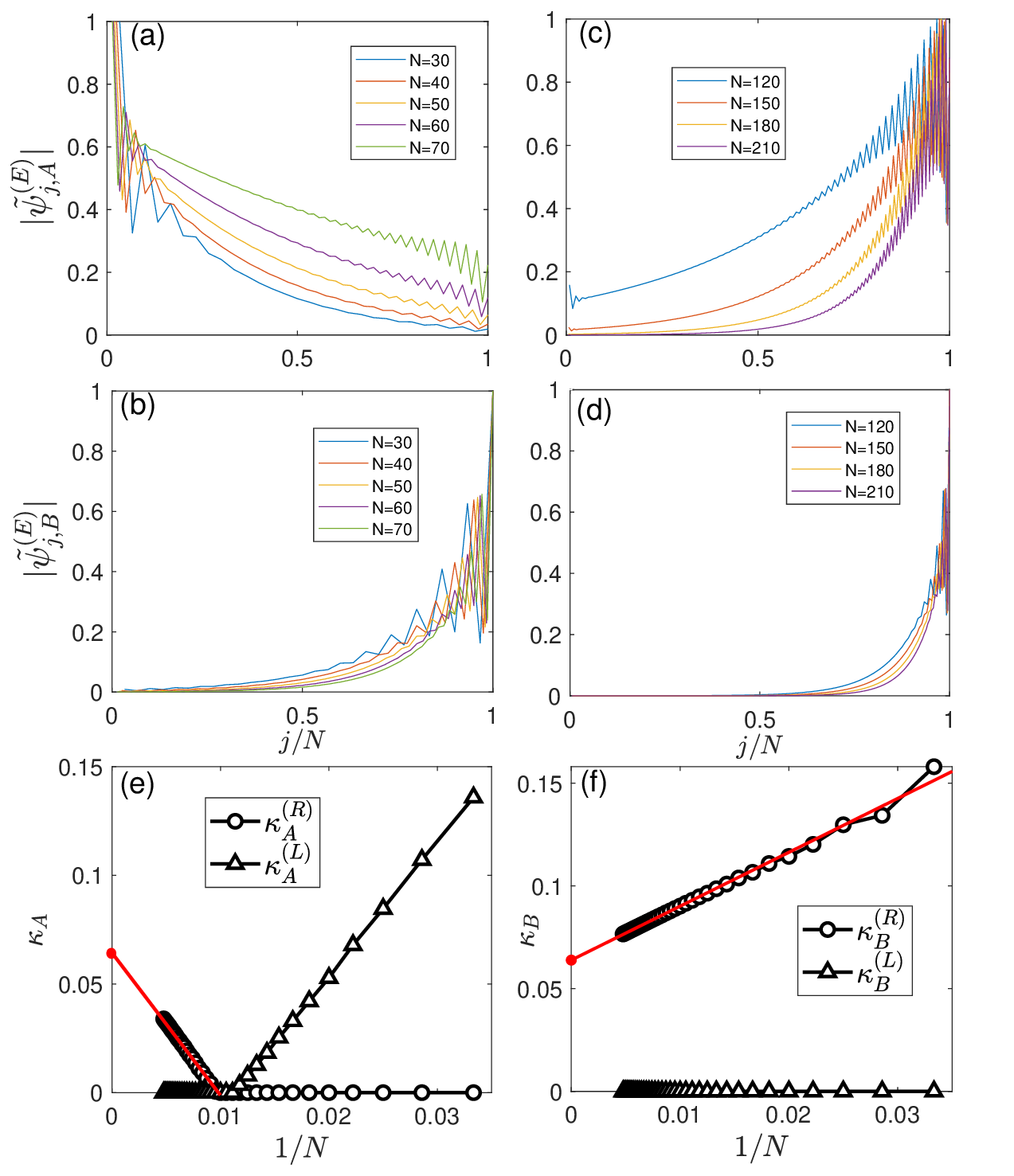}
		\caption{(a)-(d) Rescaled spatial profiles $|\tilde{\psi}_{j,\alpha}|$ of the eigenstates with the largest imaginary part of their eigenvalues as a function of $j/N$. Panels (a) and (b) correspond to system sizes $N<86$, where all states exhibit CBSE, while panels (c) and (d) show results for $N>86$, where all states exhibit unipolar NHSE. (e) Left and right inverse localization lengths, $\kappa_{A}^{(L)}$ and $\kappa_{A}^{(R)}$, for the eigenstate with the largest imaginary part as functions of $1/N$ for the wavefunctions on the $A$-chain. (f) Left and right inverse localization lengths, $\kappa_{B}^{(L)}$ and $\kappa_{B}^{(R)}$, for the eigenstate with the largest imaginary part as functions of $1/N$ for the wavefunctions on the $B$-chain. Here, $J_1=0.5$, $J_3=2$, $M=0.01$, $\delta_a=0.5$, and $\delta_{b}=0.8$.}
		\label{Fig10}
	\end{figure} 
	
	Unlike conventional critical NHSE systems, whose eigenstates exhibit the hallmark scale-free skin effect, both the concurrent bipolar skin states at small system sizes and the unipolar skin states at large system sizes in our case demonstrate a breakdown of this scale-free behavior. The scale-free skin effect is characterized by eigenstates that decay exponentially from one boundary, with the inverse localization length $\kappa$ of the wavefunctions scaling linearly with $N^{-1}$. This linear dependence means that as the system size increases, the localization length diverges, and in the thermodynamic limit ($N\to\infty$), the skin effect vanishes as $\kappa\to 0$. Consequently, when the spatial coordinate is rescaled by the system size $N$, the wavefunction profiles for different system sizes collapse onto a single universal curve, reflecting the absence of a characteristic length scale \cite{guo2021exact,li2021impurity,Guo2023}.

	Taking the case of $\delta_b>\delta_a$ as a specific example, we analyze the scaling behavior of states undergoing size-dependent transitions in our system. In Fig. \ref{Fig10}, we plot the rescaled profiles $|\tilde{\psi}_{j,\alpha}|$ of the eigenstate with the largest imaginary part of eigenvalues for various system sizes as a function of $j/N$, where $|\tilde{\psi}_{j,\alpha}| = |\psi_{j,\alpha}|/\max{|\psi_{\alpha}|}$. As shown in Figs. \ref{Fig10}(a) and \ref{Fig10}(b), which correspond to the regime where all states exhibit CBSE at small system sizes, we observe that as the system size increases, the rescaled profiles for the $A$ chain, $|\tilde{\psi}_{j,A}|$, initially localized at the left boundary, gradually migrate into bulk. Notably, this evolution occurs without the rescaled profiles collapsing onto a single universal curve as the system size increases, indicating a breakdown of scale-free behavior. In contrast, $|\tilde{\psi}_{j,B}|$ display pronounced skin like features. This deviation from the expected scale-free behavior, as reported in previous studies \cite{guo2021exact,li2021impurity,Guo2023}, highlights the distinct localization characteristics in our system. When the system size exceeds a critical value $N_c$, all states transition to exhibit unipolar NHSE. As illustrated in Figs. \ref{Fig10}(c) and \ref{Fig10}(d), further increasing the system size leads to both $|\tilde{\psi}_{j,A}|$ and $|\tilde{\psi}_{j,B}|$ becoming increasingly localized, rather than exhibiting scale-free characteristics. To further quantify the skin localization behavior, we calculate the left and right inverse localization lengths, $\kappa_{\alpha}^{(L)}$ and $\kappa_{\alpha}^{(R)}$, for the eigenstate with the largest imaginary part as functions of $1/N$. Here, the wavefunction profiles are described by $|\psi_{j,\alpha}^{(E)}|\sim e^{-\kappa_{\alpha}^{(L)}|j-1|}$ at the left boundary and $|\psi_{j,\alpha}^{(E)}|\sim e^{-\kappa_{\alpha}^{(R)}|j-N|}$ at the right boundary, with $\alpha={A,B}$, as shown in Figs. \ref{Fig10}(e) and \ref{Fig10}(f). As seen in Fig. \ref{Fig10}(e), $\kappa_A^{(L)}$ decreases linearly to zero at critical system size $N_c$, while the corresponding $\kappa_A^{(R)}$ from zero to a linear increase once the system size exceeds $N_c$, and eventually approach a finite value in the thermodynamic limit. This behavior indicates that the direction of skin localization for the $A$ chain reverses at the critical size $N_c$, which clearly deviates from scale-free behavior. In contrast, for the $B$ chain, $\kappa_B^{(L)}$ remains zero for all system sizes, whereas $\kappa_B^{(R)}$ decreases linearly to a finite value as the system size increases, as depicted in Fig. \ref{Fig10}(f). This result signifies that the wavefunction in the $B$ chain consistently exhibits skin like localization at the right boundary, but does not display scale-free characteristics.
	
	In summary, our detailed scaling analysis reveals that both the concurrent bipolar skin states at small system sizes and the unipolar skin states at large system sizes in our system fundamentally deviate from the conventional scale-free skin effect observed in typical non-Hermitian systems. Instead of universal wavefunction profiles, the localization characteristics exhibit pronounced size dependence and boundary sensitivity. These findings underscore the distinct nature of skin localization in our model and provide insights into the breakdown of scale-free behavior in non-Hermitian systems.

	\section*{Appendix C: Geometric features of spectral structure for finite interchain coupling} 
	
		\begin{figure}[b]
		\includegraphics[width=0.5\textwidth]{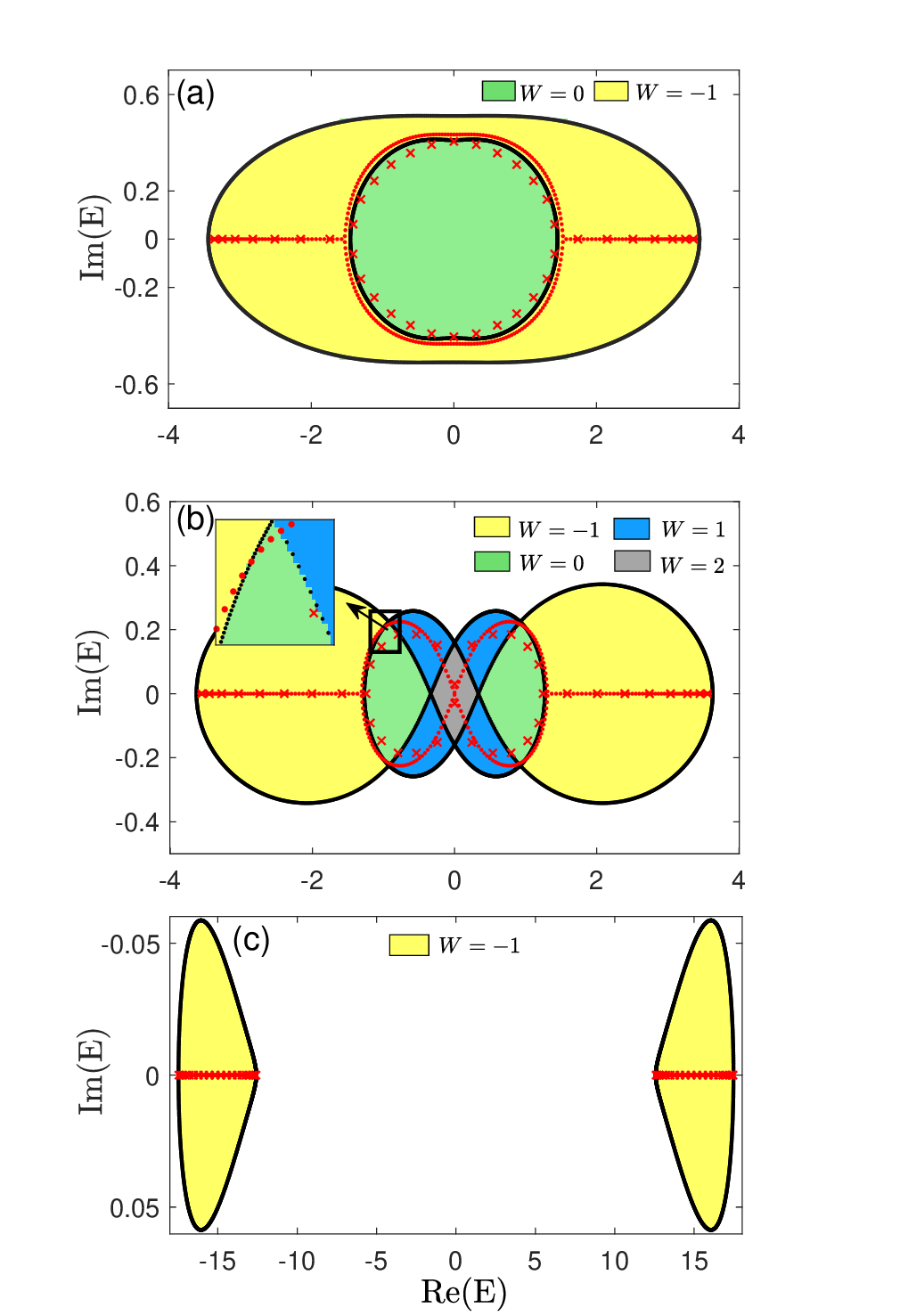}
		\caption{Energy spectra of the system under PBCs and OBCs with $J_1=0.5$, $J_3=2$, $\delta_a=0.5$, and $\delta_b=0.6$ for (a) $M=0.3$, (b) $M=0.7$, and $M=15$. The black curves indicate the PBC spectra, while red crosses and red dots represent the OBC spectra for $N=20$ and $N=150$, respectively.}
		\label{Fig12}
	\end{figure} 
	
In the main text, we focus on the regime of weak interchain coupling, where the spectral structures exhibit nested, tangent, and intersecting features for $\delta_b>\delta_a$, $\delta_b=\delta_a$, and $\delta_b<\delta_a$, respectively. As the interchain coupling amplitude $M$ increases, the spectra develop increasingly intricate geometric characteristics. To illustrate the evolution of the spectrum with varying $M$, we take the case of $\delta_b>\delta_a$ as a representative example. Figure \ref{Fig12} presents the PBC and OBC spectra for different values $M$ with $J_1=0.5$, $J_3=2$, $\delta_a=0.5$, and $\delta_b=0.6$. For $M=0.3$ [Fig. \ref{Fig12}(a)], the spectrum closely resembles that of the weak coupling limit. For small system sizes, a portion of the OBC spectrum resides within the $W=0$ region, corresponding to CBSE. As the system size exceeds a critical size $N_c$, which decreases monotonically as $M$ increases, all OBC eigenvalues migrate into the $W=-1$ region. Upon further increasing $M$, as shown in Fig. \ref{Fig12}(b) with $M=0.7$, the spectral structure undergoes a significant transformation, forming two partially overlapping figure-eight loops. For small system sizes, some OBC eigenvalues remain residing within the $W=0$ region, but as the system size becomes sufficiently large, all eigenvalues are excluded from this region. Notably, the critical system size $N_c$ exhibits a discontinuous jump due to the change in spectral topology. For sufficiently large $M$, as illustrated in Fig. \ref{Fig12}(c) with $M=15$, the PBC spectrum consists of two well-separated loops, with no $W=0$ regions enclosed by the loops. Across all system sizes, the OBC spectra are entirely real, and the corresponding eigenstates display unipolar NHSE.

	\begin{acknowledgments}
	Z.X. is supported by the NSFC (Grants No. 12375016 and 12461160324) and Beijing National Laboratory for Condensed Matter Physics (Grant No. 2023BNLCMPKF001). This work is also supported by NSF for Shanxi Province (Grant No. 1331KSC) and the National Natural Science Foundation of China (Grant No. 12474159).

	\end{acknowledgments}
	\par
	\par

\end{document}